# Improvements of Motion Estimation and Coding using Neural Networks

Raz Birman, Yoram Segal, Ofer Hadar, Senior Member, IEEE

Department of Communication Systems Engineering, BGU

Jenny Benois-Pineau, member IEEE, LABRI/ University of Bordeaux

*Abstract* – Inter-Prediction is used effectively in multiple standards, including H.264 and HEVC (also known as H.265). It leverages correlation between blocks of consecutive video frames in order to perform motion compensation and thus predict block pixel values and reduce transmission bandwidth. In order to reduce the magnitude of the transmitted Motion Vector (MV) and thus reduce bandwidth, the encoder utilizes Predicted Motion Vector (PMV), which is derived by taking the median vector of the corresponding MVs of the neighboring blocks. In this research, we propose innovative methods, based on neural networks prediction, for improving the accuracy of the calculated PMV. We begin by showing a straightforward approach of calculating the best matching PMV and signaling its neighbor block index value to the decoder while reducing the number of bits required to represent the result without adding any computation complexity. Then we use a classification Fully Connected Neural Networks (FCNN) to estimate from neighbors the PMV without requiring signaling and show the advantage of the approach when employed for high motion movies. We demonstrate the advantages using fast forward movies. However, the same improvements apply to camera streams of autonomous vehicles, drone cameras, Pan-Tilt-Zoom (PTZ) cameras, and similar applications whereas the MVs magnitudes are expected to be large. We also introduce a regression FCNN to predict the PMV. We calculate Huffman coded streams and demonstrate an order of ~34% reduction in number of bits required to transmit the best matching calculated PMV without reducing the quality, for fast forward movies with high motion.

## I. Introduction

Video coding has undergone substantial performance improvements during the last two decades. Ever increasing demand for video content, resulting from intensive use of smartphones and a wide plethora of additional rich media consumer devices coupled with expensive channels' bandwidth and higher consumer expectations, create the need to constantly reduce bandwidth requirements while retaining or improving quality. The prevailing standards used today are H.264 along with its latest improved edition H.265 (or HEVC – High Efficiency Video Coding) [1], [2] and AV1 from the Alliance of Open Media (AOMedia) [3]. The most fundamental algorithms utilized in these compression schemes are Intra-Prediction [4][5] and Inter-Prediction [1]. While Intra-Prediction takes advantage of spatial redundancies between pixel values of the same frame, Inter-Prediction leverages temporal inter-frame pixel value redundancies, e.g., the similarity between pixels in consecutive video frames.

Inter-Prediction is based on the fact that frame regions remain very close to one another or sometimes even identical between consecutive video frames. Instead of coding the original pixel block values, they are predicted from similar blocks of previous and/or future encoded frames and only the residual error is coded, thus reducing the required bandwidth. In order to reconstruct the original block pixel values at the decoder end, a Motion Vector (MV) is transmitted, which indicates the block used for the prediction. In order to further increase compression, it is desirable to reduce the value of the transmitted MV and transform the "MV-signal" with lower coding information and thus reduced coding cost. This is accomplished by predicting its value (Predicted Motion Vector or PMV) from MVs available at the encoder end and by coding only the error between the PMV and the ground truth MV which is transmitted to the decoder end. As MVs are transmitted losslessly (without quantization), the decoder recovers the current motion vector from its prediction and the decoded error. The calculated PMV is also used in Motion Estimation (ME) at the encoder, which means that for a given block, the ME algorithm searches for the best corresponding matching block in the reference frame, thus estimating the best Ground Truth block/MV (GTMV). This initial estimation can begin with the region indicated by the value of the PMV and proceed with the search for refining it. The searching algorithm is not explicitly specified in the standard and is left for the implementation. It is clear that reducing the search area will reduce calculation complexity and improve performance when searching for the GTMV. Therefore, reducing the error between the PMV and the ground truth MV will necessarily allow a better computation efficiency when searching for the GTMV at the encoder side.

In this paper, we are proposing new methods for improving the prediction of PMVs as compared to the Median vector method. Since the Median vector is not necessarily the best matching PMV, we propose to use the neighboring MV which is the best approximation of the GTMV. We examine three new different approaches for predicting the PMV. The first two methods select the best match between the 3 neighboring MV, while the third method uses the neighbors for estimating an approximated PMV – (1) signaling from encoder to encoder, with minimum extra added bandwidth; and (2) training a classification neural network to predict the best PMV from the three neighboring blocks' MVs. (3) In addition, we also propose a method, based on regression Fully Connected Neural Network (FCNN) that takes as inputs the $(x, y)$ coordinate values of the motion vectors of neighboring blocks and predicts the PMV of the block. We compare our proposed methods to the Median vector, which is the prevailing method used by the standards and demonstrate substantial improvements of Mean Squared Error (MSE) between our PMV and the GTMV, compared to the MSE between the Median vector values and the GTMV. We also calculate the entropy of the difference between our PMV and the GTMV and use Huffman coding to calculate the number of required bits for transmission. We demonstrate a promising reduction in entropy as well as the number of required bits due to the better accuracy of our proposed PMV. Reducing the number of bits required to transmit the PMV typically has an impact of up to 30% on the overall encoded stream bit rate. This improvement increases in higher values of the quantization parameter (Q), due to the higher impact that the Q has on coded DCT coefficients. The decision which blocks are estimated with intra-prediction and which ones are estimated with inter-prediction is made before quantization and therefore it is not affected by Q. While quantization is applied to DCT coefficients, it is not applied to the transmission of MV errors. Larger Q values, which reduce bit rate transmission of DCT coefficients (while decreasing quality), do not impact the number of bits required to code MV errors. So the relative impact of MV error coding on bit rate increases with higher Q values. Moreover, PMV coding impact will be higher for high motion movies, where the magnitudes of the MVs are relatively high.

Using deep learning for video coding is still an emerging research area [6][7][8]. In most research efforts, Convolutional Neural Networks (CNN) have been used to capture matching blocks features and using them for improving the predicted block, thus reducing the inter-prediction residual error [10][11][12][14]. This paper is the first research work that we are aware of, which employs neural networks and deep learning for predicting MVs.

The remainder of the paper is organized as follows. In section II we provide an overview of related research work in the area of Inter-Prediction and the commonly used PMV estimation method used by the prevailing video coding standards. In section III we present our proposed analytic method to calculate the best matching PMV. In section IV we present our proposed method of estimating the best matching MV using a classification neural network. In section V we present our proposal to estimate the PMV value using a regression neural network. In section VI we present the results of our numeric calculations. Section VII concludes the paper with a summary and future work.

## II. RELATED WORK

Various efforts have been invested in improving MV prediction accuracy and search algorithms. In [13] the authors have tackled the same challenge and have indicated the drawback of having to add signaling. They offered one selection method which is based on the content statistics, thus allowing the decoder to perform the selection without adding signaling. We use neural networks in order to accomplish that. In [15] the authors present a new technique for motion vectors prediction based on spatial and temporal prediction. The motion vector of a moving object is tracked using spatial and temporal prediction and used as a starting point for the ME searching algorithm at the encoder end. The predicted motion vector is selected from several candidate motion vectors according to the block matching criterion. Experiments show that this spatial-temporal prediction reduces the number of computations performed by the motion search algorithm by 30% for MPEG2 encoding and by 40% for H.263 encoding. The Median method has typically yielded sufficiently accurate results of the GTMV for coding purposes; therefore the majority of the research efforts have been invested in improving the efficiency of the motion estimation itself. In [16] a MV prediction method is presented. It is a Prediction Search Algorithm (PSA) for block motion estimation. The proposed method utilizes a linear combination of the motion vectors of the three adjacent blocks to obtain a predicted motion vector, namely, the initial search point. Simulation results show that the proposed PSA is better than the three-step search algorithm [17] and the four-step search algorithm [18] in terms of MSE with smaller computational workload. To improve the accuracy of the fast Block Matching Algorithms (BMAs), in [20], a new adaptive motion tracking search algorithm is proposed. Based on the spatial correlation of motion blocks, a predicted starting search point, which reflects the motion trend of the current block, is adaptively chosen. Experimental results show that the proposed algorithm enhances the accuracy of the fast center-biased Block-Matching-Algorithms (BMAs), such as the new three-step search [9], the four-step search [18], and the block-based gradient descent search [19], as well as reduces their computational cost. As in [17], L. Luo. *et al.*, [21] propose a new prediction search algorithm for block motion estimation utilizing the linear weighing of the MVs of the 3 adjacent blocks. In [22] E. Kaminsky and O. Hadar propose a method for effectively analyzing and selecting the most suitable motion estimation algorithm. All these methods use conventional prediction approaches such as least square estimation of weights in a linear combination of neighboring vectors, Median prediction and so on. Due to substantial improvements in compression efficiency, the contribution of PMV error coding is becoming more attractive as means to obtain further bit rate reduction while retaining the same quality (since it is not quantized).

In recent years there have been efforts to harness the power of Neural Networks for improving predictions for video coding. More efforts have been invested in the improvement of Intra-Prediction [9][23][24][25][26]. However, there has also been some research for improving motion compensation. The primary focus of these research papers was on motion compensation at the single pixel level, which corresponds to optical flow. Thus, the authors of [27] use Convolutional Neural Networks (CNN) for predicting a heat map optical flow from consecutive video frames.

*A. Calculation of Predicted Motion Vector (PMV)*

The prevailing algorithms developed for video coding standards calculate the PMV by taking the Median vector of the neighboring blocks' MVs. This can be accomplished assuming that neighboring blocks have been predicted with Motion Estimation. However, some blocks are predicted with Intra-Prediction and not with Inter-Prediction. So in some cases there are less than 3 Motion Estimated neighbors. Therefore, the prediction applies only to a subset of the frame blocks. The Median vector is calculated for blocks that have three, two and one neighboring Motion Estimated blocks respectively.

*B. Primary benefits of more accurately predicted PMVs*

A more accurate prediction of the PMVs can improve two different aspects of the compression algorithm:

(1) more accurately predicted PMVs are expected to produce a lower error compared to the ground truth MV, thus reducing the number of bits required to represent the difference. Some caution should be exercised due to Huffman coding [29]. The important criteria for determining bit rate, when using Huffman coding, is the probability distribution of the coded values, which can be expressed by the normalized histogram of the predicted PMV residual values. If the motion vectors are accurately predicted, then the low error values will be more probable thus the number of bits required in entropic Hoffman coding will be lower.

(2) the value of the PMV is usually used for motion compensation calculation at the encoder end, which points out the block with the largest pixels similarity to the predicted block. When the PMV is closer to the ground truth vector, it is possible to produce more efficient searching regions as well as increase the accuracy of finding the best matching vector, thus improving computation efficiency as well as reducing residual error.

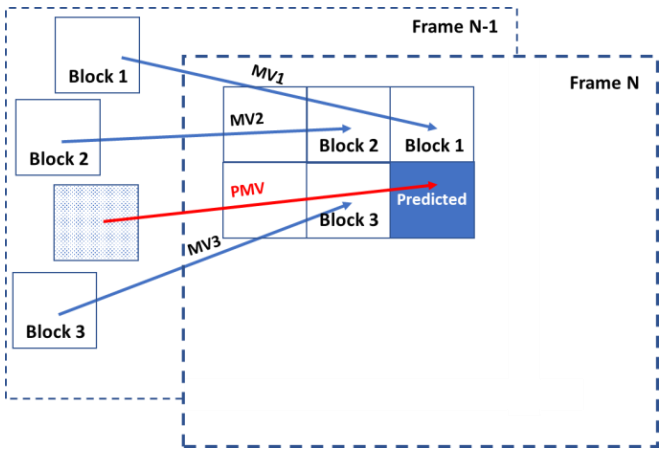

*Fig. 1. Illustration of Inter-Prediction calculated PMV from 3 neighboring Motion- Estimated blocks*

*C. Evaluation criteria*

The common method to assess compression performance improvement results is Rate-Distortion (RD) curves. RD curves present the relation between bit rates and quality as it is calculated for different values of the quantization parameter (Q). RD curves provide the overall compression performance assessment criteria that encompasses spatial as well as temporal prediction algorithms. This research paper deals with improving the compression corresponding to encoding motion compensation MV errors. Quantization is not used for transmitting errors of PMVs. Moreover, as explained earlier, their impact changes with Q due to the dominant impact it has on the encoding of DCT coefficients. In order to isolate the contribution of our proposed new algorithms and evaluate their performance, we have determined to use entropy and Huffman coding. When using them as criteria, we can focus on motion estimation performance and compare different algorithms to improve it.

### III. Calculating Best PMV

To predict the PMV of the current block, the Inter-Prediction schemes use motion vectors of three surrounding blocks as illustrated in **Fig. 1**. The standard calculation of the PMV derives the Median vector of the MVs of surrounding motion estimated blocks as indicated in equation (1).

$$\overrightarrow{PMV_i} = \overrightarrow{Median_i} = Median$$
$$\overrightarrow{MV_i} = Median + \overrightarrow{Err_i} \quad \{i = 1 \dots k\} \quad (1)$$

Where k is the number of motion compensated predicted blocks in the frame.

The median is calculated for the $x$ as well as for the $y$ component. The formula is depicted in equation (2).

$$x_{PMV} = Median\{x_{MV1}, x_{MV2}, x_{MV3}\}$$
$$y_{PMV} = Median\{y_{MV1}, y_{MV2}, y_{MV3}\} \quad (2)$$

Whereas MV1, MV2 and MV3 represent the three surrounding MVs, as illustrated in **Fig. 1**. However, as it turns out, the median is not necessarily the best approximation out of the three. Let $MV_{i,j} = \{x_{i,j}, y_{i,j}\}$ be the GTMV of block in frame position $i, j$. Let $MV_{i-1,j} = \{x_{i-1,j}, y_{i-1,j}\}$, $MV_{i-1,j-1} = \{x_{i-1,j-1}, y_{i-1,j-1}\}$, and $MV_{i,j-1} = \{x_{i,j-1}, y_{i,j-1}\}$ be the MVs of the 3 neighboring blocks – block on the right, block on the top corner, and block on the top respectively. Let the median of the neighboring MVs in the $x$ and in the $y$ coordinates respectively be represented by equation (3):

$$Median_{X_{i,j}} = Median\{x_{i-1,j}, x_{i-1,j-1}, x_{i,j-1}\}$$
$$Median_{Y_{i,j}} = Median\{y_{i-1,j}, y_{i-1,j-1}, y_{i,j-1}\} \quad (3)$$

Let the delta between the median and the GTMV of block $i, j$ and that median be as describe in equation (4):

$$\Delta_{median\_x_{i,j}} = Median_{X_{i,j}} - x_{i,j} \quad (4)$$

$$\Delta_{median\_y_{i,j}} = Median_{Y_{i,j}} - y_{i,j}$$

Instead of the median we propose an analytic method to find the neighbor MV with the lowest difference from GTMV along the $x$ and the $y$ coordinates respectively (see equation (5):

$$\Delta_{Best_{x_{i,j}}} = minimum\{|x_{i-1,j} - x_{i,j}|, |x_{i-1,j-1} - x_{i,j}|, |x_{i,j-1} - x_{i,j}|\}$$

$$\Delta_{Best_{y_{i,j}}} = minimum\{|y_{i-1,j} - y_{i,j}|, |y_{i-1,j-1} - y_{i,j}|, |y_{i,j-1} - y_{i,j}|\} \quad (5)$$

The Mean Square Error (MSE) between the median and the GTMV for the frame will be as describe in equation (6):

$$MSE_{median\_x} = \frac{1}{N}\sum_{i,j}\Delta_{median\_x_{i,j}}^2$$

$$MSE_{median\_y} = \frac{1}{N}\sum_{i,j}\Delta_{median\_y_{i,j}}^2 \quad (6)$$

Where $N$ represents the number of blocks.

The MSE between the GTMV and the lowest difference neighbor is defined in equation (7):

$$MSE_{Best\_x} = \frac{1}{N}\sum_{i,j}\Delta_{Best\_x_{i,j}}^2$$

$$MSE_{Best\_y} = \frac{1}{N}\sum_{i,j}\Delta_{Best\_y_{i,j}}^2 \quad (7)$$

By definition, as indicated in equation (8):

$$MSE_{Best\_x} \leq MSE_{median\_x}$$

$$MSE_{Best\_y} \leq MSE_{median\_y} \quad (8)$$

Therefore, our best PMV is derived by finding the MV with the lowest MSE when compared to the GTMV. The same MV will also provide the best coding efficiency since it will also yield lower entropy than the median, which in turn results in less required coding bits. In the results section below, we have shown the entropy as well as performed Huffman coding to calculate the number of bits required to code the residual error between the GTMV and the best PMV. The drawback of this approach is that we have to add signaling to the code stream in order to notify the decoder of the selected PMV per each and every motion compensated block. However, since the median can be selected as default, we only need to signal in the case of selecting one of the other neighboring MVs, therefore we can reduce the required signaling to average less than one bit per block. Since we deal with blocks, which have three motion compensated neighbors, we need to select one out of the three. The decoder can calculate the median, so we only need to signal whether the PMV is larger or smaller than the median. Such signaling can be accomplished with one extra bit for non-median blocks.

IV. USING CLASSIFICATION NN TO PREDICT BEST PMV

Signaling can be altogether avoided by training a classification neural network to predict the best PMV. The network is trained once on a dataset of blocks that is retrieved from multiple movies. Once trained, all network weights are set and can be used at runtime efficiently by the encoder and the decoder to perform the prediction from the MVs of neighboring blocks. The inputs of the network will be the 3 coordinate pairs of the neighboring MVs and the argument of the median, which is the index of the corresponding MV, as indicated in equation (9).

$$input_{i,j} = \{x_{i-1,j}, y_{i-1,j}, x_{i-1,j-1}, y_{i-1,j-1}, x_{i,j-1}, y_{i,j-1}, Arg_{median_x}, Arg_{median_y}\} \quad (9)$$

Classification network Softmax layer outputs three probabilities and the highest probable MV is selected as the best PMV. We provide the network as inputs also the arguments of the median MV indexes to assist it to distinguish between close classification probabilities. This is since the probabilities may be close in some cases and since we know that the median is, in some cases, the best PMV.

The output neuron of the network with the highest probability of classified as best PMV is indicated in equation (10) for the $x$ and $y$ coordinates respectively, where $\emptyset_{x_{i,j}}$ and $\emptyset_{y_{i,j}}$ represent the index of the neuron with the highest probability form block $i,j$ in the $x$ and $y$ coordinates respectively.

$$\phi_{x_{i,j}}(input) \approx Argmin\{(x_{i-1,j} - x_{i,j}), (x_{i-1,j-1} - x_{i,j}), (x_{i,j-1} - x_{i,j})\}$$

$$\phi_{y_{i,j}}(input) \approx Argmin\{(y_{i-1,j} - y_{i,j}), (y_{i-1,j-1} - y_{i,j}), (y_{i,j-1} - y_{i,j})\} \quad (10)$$

The classification network that we have used has an input layer, 5 hidden layers with 8 neurons each, and a softmax layer with 3 outputs. The input is specified in equation (9). The outputs are 3 probabilities corresponding to the 3 neighboring MVs. The highest probability output indicates the predicted best PMV, as indicated in equation (11).

$$Arg_{PMV\_x} = Argmax\{Pr_{X_{MV1}}, Pr_{X_{MV2}}, Pr_{X_{MV3}}\}$$

$$Arg_{PMV\_y} = Argmax\{Pr_{Y_{MV1}}, Pr_{Y_{MV2}}, Pr_{Y_{MV3}}\} \quad (11)$$

Since MV values can be positive and negative, and since the argument coordinates of the median vector are different by nature from the continuum of the MV values, we have normalized all MV coordinates to the range of -0.8 – 0.8 and have mapped the 3 possible median argument values to 0.85, 0.90, and 0.95 respectively, as indicated in equation (12). Using this mapping scheme, we have accomplished an input range of -0.8 – 0.95 and clearly and consistently distinguished the median coordinates from the $(x, y)$ coordinates of the input MVs, thus improving the network learning capacity.

$$Input_{MV_{x_{i,j}}} = 0.8 * MV_{x_{i,j}} / \max_{\forall i,j}\{MV_{x_{i,j}}\}$$

$$Input_{MV_{y_{i,j}}} = 0.8 * MV_{y_{i,j}} / \max_{\forall i,j}\{MV_{y_{i,j}}\}$$

$$Input_{Arg_{median_x}} = Arg_{median_x} * 0.05 + 0.85 \tag{12}$$

$$Input_{Arg_{median_y}} = Arg_{median_y} * 0.05 + 0.85$$

We further used the $tanh$ activation function, which has a suitable dynamic range of $\{-1 : 1\}$. The classification network is illustrated in **Fig. 2.**

The classification accuracy accomplished with 3 neighboring MVs varies between 70% - 80%, depending on dominant objects' movements between video frames. We have learned that it is less likely to obtain sufficient improvement of the PMV when testing movies with relatively small movements, e.g., the small magnitude of MVs. In order to further substantiate this observation we have tested fast forward (FFW) movies.

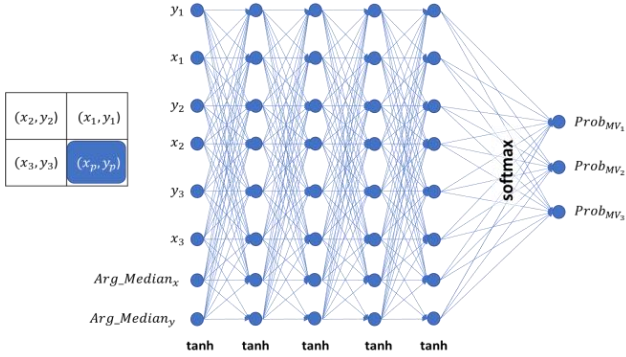

*Fig. 2: Classification network to find the best PMV $(x_p, y_p)$*

FFW movies are created by skipping frames. The resulting MVs between 2 consecutive frames are calculated as an aggregation of the MVs between the skipped frames, as described in [30]. Therefore, the resulting MVs are of higher magnitudes. We further substantiated this observation by calculating MV statistics of our datasets (see section VI-D below). We show 35% - 66% improvement of MSE and corresponding 3% - 7% saving of required coding bits when running our algorithms on selected FFW movies.

## V. USING REGRESSION NN TO PREDICT THE PMV

An additional method that we propose is predicting the PMV using trained regression FCNN. We have used a FCNN that includes 1 to 5 hidden layers. The network is fed with 6 numbers, representing the $\{x; y\}$ values of the surrounding block MVs and outputs a value, representing the $x$ or $y$ value of the PMV of the current block. The network architecture (with an example of 3 hidden layers) is depicted in **Fig. 3**. It works as a regressor using Euclidean loss function which, in our case, is a mean squared error between predicted and ground-truth motion vectors resulting from the motion estimation algorithm (see equation (13)). Two different networks of identical architecture were trained to predict the $x$ and $y$ values of the PMV respectively.

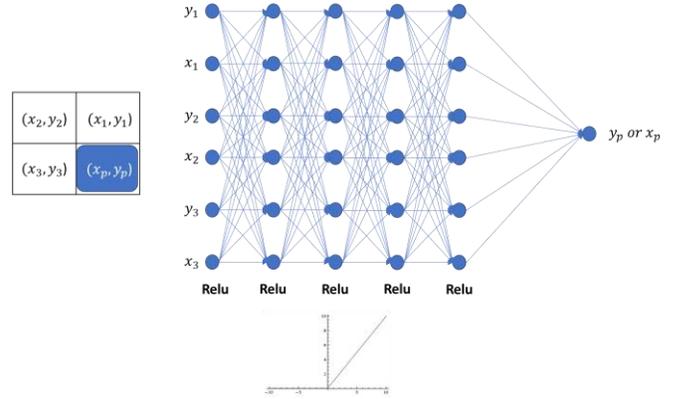

*Fig. 3: FCNN architecture used to predict the value of the PMV $(x_p, y_p)$*

## VI. RESULTS

### A. Dataset

The dataset used for training is extracted from the Table Tennis movies of UCF101 videos dataset [28]. Here strong and irregular motions are observed by visual inspection of selected videos. An illustration of 2 arbitrary frames from the dataset is provided in **Fig. 4**. We scanned the data URLs and extracted six different datasets from multiple randomly selected Table Tennis movies. We used as many video files as necessary to satisfy the specified number of blocks that we used for training and for testing respectively. Motion-compensated blocks with zero movements were ignored when compiling the blocks' dataset. In all cases we used one dataset for training and tested with the remaining 5 datasets while averaging the performance results. An illustration of the MVs superimposed on the frame is provided in **Fig. 5.** The videos we worked with have been analyzed to have blocks of sizes: 16x16, 8x8, 8x16, 16x8 due to the target compression standard (H.264). These 4 different blocks create a variety of $4^3 = 64$ permutations of neighboring cases. A subset of the cases, that demonstrates the considered neighbors is depicted in *Fig. 6*. To assess the accuracy of the proposed prediction scheme, we calculate Mean Squared Error (MSE) of equation (13) over the full dataset.

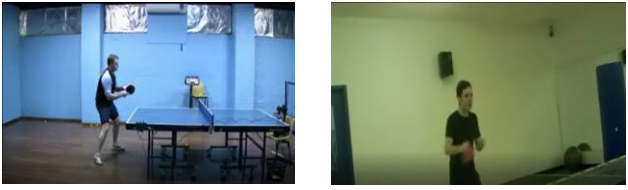

*Fig. 4: Sample frames from the UCF101 Table Tennis Dataset*

We have used FFMPEG to extract MVs from the movies. In videos with strong object motions, such as sports movies with large body parts movements, the MVs distribution is dispersed, as illustrated in an example from UCF101 table tennis sequences in **Fig. 4** with motion vectors illustrated in **Fig. 5**.

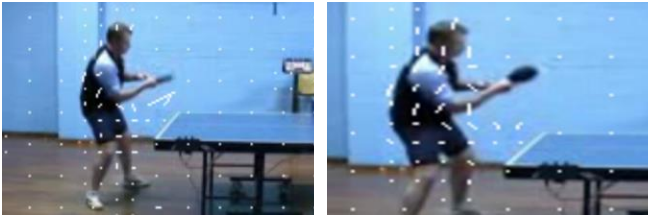

*Fig. 5: Sample partial frames with superimposed MVs*

$$MSE = \frac{1}{N}\sum_N \{(\hat{x}-x)^2 + (\hat{y}-y)^2\} \quad (13)$$

Whereas:

N is the number of samples in the dataset

x̂, ŷ are the predicted values of the block's MV (PMV)

x, y are the ground truth values of the block's MV (GTMV)

Using the 'Table Tennis' videos subset of the UCF101 dataset [28] we have extracted 50,000 blocks that satisfy a neighboring criterion of having 3, 2 or 1 motion compensated blocks respectively. All remaining blocks are estimated with Intra-Prediction and therefore not relevant to our algorithms. Since not all blocks have neighbors predicted with motion compensation, we have divided the dataset into three different categories according to the following neighboring criteria:

1. Blocks with 3 motion compensation neighbors
2. Blocks with 2 motion compensation neighbors
3. Blocks with 1 motion compensation neighbor

For testing we have extracted a sample of 2,000 blocks from the same dataset, while ensuring that testing blocks and training blocks are extracted from different movies.

### B. Calculating best PMV

At the encoder side, we have all the information necessary to calculate the best matching neighboring MV, which is the closet in value to the GTMV. We have performed these calculations and compared them to the median MV. We compared MSE relative to the GTMV, entropy of the difference from the GTMV (which is an indication of coding bits requirement) and the actual number of bits, which are calculated using Huffman coding of all PMVs over the datasets. **Table 1** and **Table 2** provide the calculated MSE, Entropy and number of required coding bits for the $\Delta x$ and for the $\Delta y$ coordinates respectively. The average savings indicated by these numbers are consistent for both coordinates and yields 75% improvement of MSE, 42% improvement of the entropy, 35% reduction of bits required to code the $MV_x$ component and 31% improvement of bits required to code the $MV_y$ component. When averaging over the datasets, these improvements have very small variation, in the order of ±0.005%, when considering the standard deviation of the results. The average savings percentages are depicted in **Table 3**.

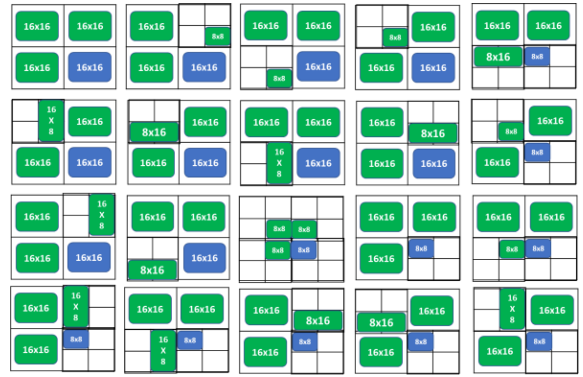

*Fig. 6: Subset of the possible block neighbors used to calculate the MVs*

**Table 1:** MSE, Entropy and number of Bits of the Median compared to the best PMV with respect to the GTMV. Calculated for Δx using 5 different datasets and averaged

|  | $Median_x$ | | | Best $PMV_x$ | | |
|---|---|---|---|---|---|---|
|  | MSE | Entropy | # Bits | MSE | Entropy | # Bits |
| Dataset1 | 8.9888 | 1.8769 | 192,441 | 2.2057 | 1.0883 | 123,665 |
| Dataset2 | 8.8184 | 1.882 | 206,857 | 2.1678 | 1.0868 | 132,585 |
| Dataset3 | 8.3882 | 1.8691 | 206,049 | 2.1195 | 1.0864 | 132,959 |
| Dataset4 | 8.5616 | 1.8769 | 203,521 | 2.1137 | 1.0803 | 130,275 |
| Dataset5 | 8.3929 | 1.8224 | 203,674 | 2.1761 | 1.0584 | 132,588 |
| Average± Std Dev | 8.63± 0.27 | 1.87± 0.02 | 202,508± 5,813 | 2.1566± 0.04 | 1.08± 0.01 | 130,414± 3,921 |

*Table 2: MSE, Entropy and number of Bits of the Median compared to the best PMV with respect to the GTMV. Calculated for Δy using 5 different datasets and averaged*

|  | $Median_y$ | | | Best $PMV_y$ | | |
|---|---|---|---|---|---|---|
|  | MSE | Entropy | # Bits | MSE | Entropy | # Bits |
| Dataset1 | 5.4035 | 1.5366 | 159,700 | 1.3144 | 0.8909 | 110,154 |
| Dataset2 | 5.1995 | 1.5099 | 168,999 | 1.2639 | 0.8735 | 117,036 |
| Dataset3 | 5.095 | 1.535 | 171,588 | 1.2382 | 0.888 | 118,356 |
| Dataset4 | 5.2757 | 1.5268 | 168,131 | 1.3149 | 0.8847 | 116,283 |
| Dataset5 | 4.8822 | 1.4789 | 168,476 | 1.1945 | 0.8463 | 117,063 |
| Average± Std Dev | 5.1712± 0.2 | 1.5174± 0.02 | 167,379± 4,502 | 1.2652± 0.05 | 0.8767± 0.02 | 115,778± 3,231 |

These improvements are substantial. However, they do not include the signaling bits required to indicate to the decoder which is the best PMV. The results of including the signaling bits for the cases when the median is not the selected PMV, are

Inter-Prediction with Deep Learning

presented in **Table 4**. The results show an average reduction of 23% and 18% of bits required to code the $MV_x$ component the $MV_y$ component respectively.

*Table 3: % improvement of MSE, Entropy and number of Bits with respect to the Median vector. Calculated for $\Delta x$ and for $\Delta y$, using 5 different datasets and averaged*

|          | $1 - \nabla_{Best_x}/\nabla_{Median\_x}$ | | | $1 - \nabla_{Best_y}/\nabla_{Median\_y}$ | | |
|----------|------|---------|--------|------|---------|--------|
|          | MSE  | Entropy | # Bits | MSE  | Entropy | # Bits |
| Dataset1 | 75%  | 42%     | 36%    | 76%  | 42%     | 31%    |
| Dataset2 | 75%  | 42%     | 36%    | 76%  | 42%     | 31%    |
| Dataset3 | 75%  | 42%     | 35%    | 76%  | 42%     | 31%    |
| Dataset4 | 75%  | 42%     | 36%    | 75%  | 42%     | 31%    |
| Dataset5 | 74%  | 42%     | 35%    | 76%  | 43%     | 31%    |
| Average  | 75%±0.004% | 42%±0% | 36%±0.005% | 76%±0.004% | 42%±0.004% | 31%±0% |

*Table 4: Saving of required coding bits, compared to the Median vector, when using the best PMV and adding signaling*

| # Bits   | $x$ | | | $y$ | | |
|----------|-----------------|----------------|----------|-----------------|----------------|----------|
|          | $\nabla_{Median}$ | $\nabla_{Best}$ | % Saving | $\nabla_{Median}$ | $\nabla_{Best}$ | % Saving |
| Dataset1 | 153,401 | 117,667 | 23% | 105,870 | 89,105  | 16% |
| Dataset2 | 146,390 | 116,712 | 20% | 125,586 | 103,100 | 18% |
| Dataset3 | 160,488 | 123,120 | 23% | 117,159 | 96,127  | 18% |
| Dataset4 | 170,982 | 128,340 | 25% | 123,469 | 100,573 | 19% |
| Dataset5 | 150,068 | 115,673 | 23% | 112,512 | 92,764  | 18% |
| Average  |         |         | 23%±2% |         |         | 18%±1% |

### C. Using a classification network to predict best PMV

We can save the signaling by training a classification neural network to predict the best PMV from the 3 neighboring motion compensated blocks, as described in section IV. We have trained two different networks with the architecture depicted in **Fig. 2** – one for the $PMV_x$ component and second for the $PMV_y$ component. The training accuracy that is obtained is 70.4% for $PMV_x$ and 79% for $PMV_y$ respectively. The network convergence graphs with training epochs are depicted in **Fig. 7** and **Fig. 8** for the $PMV_x$ component and the $PMV_y$ component respectively. The training data is split with 30% validation set and the accuracy is calculated on the validation set during training. The MSE, Entropy and number of required coding bits were calculated for the $PMV_x$ and the $PMV_y$ components.

The results are provided in **Table 5** and **Table 6** respectively. As can be seen in **Table 7**, the classification network accuracy is insufficient to obtain improvement. The average degradation of MSE is of 16% and 21%, average degradation in entropy is 7% and 9% for the $MV_x$ and the $MV_y$ components respectively. The average degradation in the number of bits required for coding is 6% and 7% for $MV_x$ and the $MV_y$ respectively.

Nevertheless, we have observed that the degradation depends on the magnitude of the MVs. In order to further substantiate this observation we have applied the classification network to FFW movies.

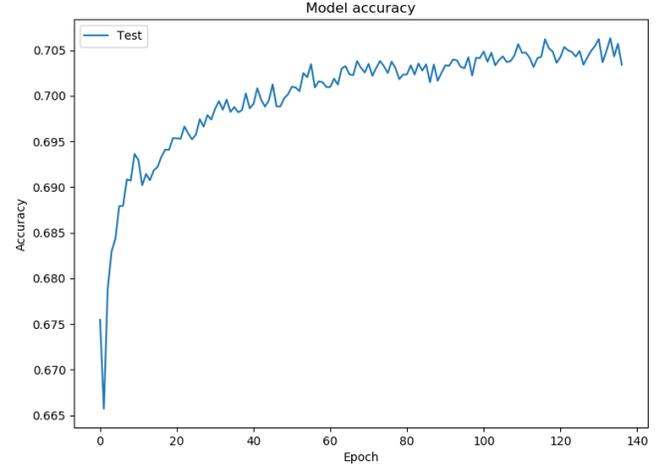

*Fig. 7: Classification network convergence when training for $PMV_x$ component*

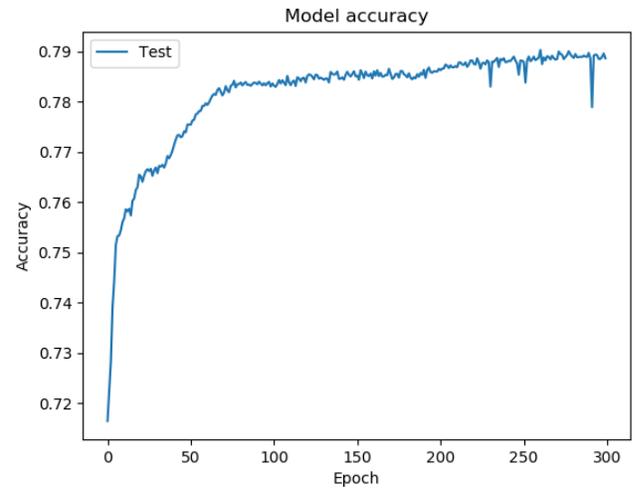

*Fig. 8: Classification network convergence when training for $PMV_y$ component*

**Table 5:** MSE, Entropy and number of Bits of the Median vector compared to the classification network predicted best PMV with respect to the GTMV. Calculated for $\Delta x$ using 5 different datasets and averaged

|          | $Median_x$ | | | Best $PMV_x$ (Classification) | | |
|----------|--------|---------|--------|--------|---------|--------|
|          | MSE    | Entropy | # Bits | MSE    | Entropy | # Bits |
| Dataset1 | 4.3528 | 1.707   | 22,222 | 5.3815 | 1.7781  | 23,066 |
| Dataset2 | 4.7668 | 1.3244  | 19,835 | 6.4904 | 1.5478  | 22,381 |
| Dataset3 | 8.304  | 1.7887  | 20,746 | 8.5806 | 1.8989  | 21,982 |
| Dataset4 | 8.0678 | 1.8207  | 19,301 | 8.5573 | 1.9172  | 20,275 |
| Dataset5 | 5.3619 | 1.7695  | 19,695 | 5.8626 | 1.7921  | 19,832 |
| Average± Std Dev | 6.1707±1.88 | 1.6821±0.2 | 20,360±1,168 | 6.9745±1.51 | 1.7868±0.15 | 21,507±1,391 |

### D. Application for Fast Forward movies

Since FFW movies are created by skipping frames and since the final movie MVs are an aggregation of the MVs between the individual skipped frames [30], the magnitude of MVs in FFW movies is much larger than that of regular movies, and therefore our classification network efficiency is sufficient to obtain improvement.

Inter-Prediction with Deep Learning

*Table 6: MSE, Entropy and number of Bits of the Median vector compared to the classification network predicted best PMV with respect to the GTMV. Calculated for $\Delta y$ using 5 different datasets and averaged*

|  | $Median_y$ | | | Best $PMV_y$ (Classification) | | |
|---|---|---|---|---|---|---|
|  | MSE | Entropy | # Bits | MSE | Entropy | # Bits |
| Dataset1 | 2.6193 | 1.4594 | 19,379 | 3.2857 | 1.5642 | 20,516 |
| Dataset2 | 1.1909 | 0.9895 | 16,232 | 1.5902 | 1.1639 | 17,989 |
| Dataset3 | 4.517 | 1.562 | 18,348 | 4.6598 | 1.6541 | 19,308 |
| Dataset4 | 4.3742 | 1.5877 | 17,084 | 4.9874 | 1.6801 | 17,903 |
| Dataset5 | 3.9654 | 1.3801 | 15,922 | 5.1618 | 1.5087 | 17,032 |
| Average± Std Dev | 3.3333± 1.41 | 1.3957± 0.24 | 17,393± 1,455 | 3.937± 1.5 | 1.5142± 0.21 | 18,550± 1,367 |

*Table 7: % degradation of MSE, Entropy and number of Bits with respect to the Median vector. Calculated for $\Delta x$ and for $\Delta y$, using 5 different datasets and averaged*

|  | $1 - \nabla_{Best_x}/\nabla_{Median\_x}$ (Classification) | | | $1 - \nabla_{Best_y}/\nabla_{Median\_y}$ (Classification) | | |
|---|---|---|---|---|---|---|
|  | MSE | Entropy | # Bits | MSE | Entropy | # Bits |
| Dataset1 | -24% | -4% | -4% | -25% | -7% | -6% |
| Dataset2 | -36% | -17% | -13% | -34% | -18% | -11% |
| Dataset3 | -3% | -6% | -6% | -3% | -6% | -5% |
| Dataset4 | -6% | -5% | -5% | -14% | -6% | -5% |
| Dataset5 | -9% | -1% | -1% | -30% | -9% | -7% |
| Average | -16%± 14% | -7%± 6% | -6%± 4% | -21%± 13% | -9%± 5% | -7%± 2% |

We used 2 different cars traffic FFW movies. One frame of each movie is depicted in **Fig. 9**. We compared the MV statistics of blocks' subsets extracted from these 2 movies to those of our UCF101 Table Tennis dataset. The results of average MV magnitudes as well as standard deviations over 50,000 sampled blocks are depicted in **Table 8**. As can be seen, the standard deviation of the MV of the FFW movies is much larger than that of the Table Tennis movies, thus indicating much larger MV magnitudes.

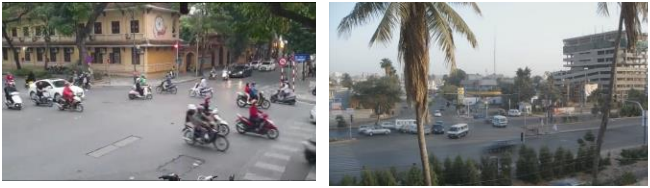

*Fig. 9: Sample of one frame from two different FFW movies*

*Table 8: MV statistics for datasets extracted from UCF101 Table Tennis movies compared to two FFW movies*

|  | $MV_x$ | | $MV_y$ | |
|---|---|---|---|---|
|  | Average | Std Dev | Average | Std Dev |
| Table Tennis Movies | -0.0394± 0.079 | 3.1092± 0.37 | 0.1748± 0.25 | 2.438± 0.21 |
| FFW Movies | 3.1377± 3.23 | 22.7361± 10.41 | -0.6803± 0.49 | 4.4247± 1.76 |

We trained the classification neural network of **Fig. 2** for these movies and calculated the resulting MSE, entropy and number of required bits compared to the median. The results are depicted in **Table 9** and in **Table 10** for $PMV_x$ and $PMV_y$ respectively.

The accuracy that was obtained during training was 84% and 80% for $MV_x$ and $MV_y$ respectively. The improvement results of the MSE, entropy and number of coding bits are depicted in **Table 11**. As can be seen, there is a large variation of the results in the $\Delta x$ and the $\Delta y$ as well as between the two movies. This is due to the content, which has different distributions of MVs in these dimensions.

*Table 9: MSE, Entropy and number of Bits of the Median vector compared to the classification network predicted best PMV with respect to the GTMV. Calculated for $\Delta x$ using 2 FFW movies.*

|  | $Median_x$ | | | Best $PMV_x$ (Classification) | | |
|---|---|---|---|---|---|---|
|  | MSE | Entropy | # Bits | MSE | Entropy | # Bits |
| Movie 1 | 87.3705 | 1.821 | 5,617 | 37.4915 | 1.678 | 5,156 |
| Movie 2 | 600.5415 | 2.3094 | 7,374 | 144.8425 | 2.1522 | 6,580 |

Repeating the calculation for FFW movies with the analytically derived best PMV of section *B* above, we obtain improvements of 20% and 11% for $PMV_x$ and $PMV_y$ respectively.

*Table 10: MSE, Entropy and number of Bits of the Median vector compared to the classification network predicted best PMV with respect to the GTMV. Calculated for $\Delta y$ using 2 FFW movies.*

|  | $Median_y$ | | | Best $PMV_y$ (Classification) | | |
|---|---|---|---|---|---|---|
|  | MSE | Entropy | # Bits | MSE | Entropy | # Bits |
| Movie 1 | 4.7835 | 1.096 | 3,779 | 4.6185 | 1.096 | 3,714 |
| Movie 2 | 26.346 | 1.6281 | 5,197 | 8.551 | 1.5354 | 4,720 |

*Table 11: % improvement of MSE, Entropy and number of Bits with respect to the Median vector. Calculated for $\Delta x$ and for $\Delta y$, Calculated for 2 FFW movies.*

|  | $1 - \nabla_{Best_x}/\nabla_{Median\_x}$ (Classification) | | | $1 - \nabla_{Best_y}/\nabla_{Median\_y}$ (Classification) | | |
|---|---|---|---|---|---|---|
|  | MSE | Entropy | # Bits | MSE | Entropy | # Bits |
| Movie 1 | 57% | 8% | 8% | 3% | 0% | 2% |
| Movie 2 | 76% | 7% | 11% | 68% | 6% | 9% |

### E. Using regression to predict the PMV

We have implemented a regression network of the architecture depicted in **Fig. 3** to predict the value of the PMV based on the MVs of the neighboring blocks. The regression network learns to predict a PMV value. This PMV is not necessarily identical to one of the neighboring block's MVs, which were predicted by the classification network of the previous method. We trained the network with 50,000 blocks with a validation split of 30%. The loss function minimizes the MSE between the predicted PMV and the ground truth MV for each one of the PMV components separately. The regression network typically converges according to that criteria before reaching 50 epochs of the complete dataset size (no batching used in training). We trained two separate networks, for predicting $x$ and $y$ of the PMV respectively, for each case using the extracted MVs dataset. The participating movies were selected randomly from within the dataset. The test was run on a dataset of 2,000 blocks, also selected from randomly selected movies from the same dataset, ensuring that they are always different from the ones used for training. The test was run 5 times for different randomly selected datasets. The results for blocks with 3 motion-compensated neighboring blocks and with 2 neighboring blocks are summarized in **Table 12** and **Table 13** respectively. As can be seen, the prediction results for 3 motion compensated neighbors are ~20% better in terms of MSE than those of the Median vector. The trained network results for blocks with 2 motion-compensated vectors, whereas the median is actually the average of the two, are not as good but still better compared to the Median vector.

*Table 12: MSE of regression Network-Predicted versus Median-Predicted for blocks with 3 motion compensated neighbors*

| 3 MVs | $1 - \nabla_{PMV_x}/\nabla_{Median\_x}$ (Regression) | | | $1 - \nabla_{PMV_y}/\nabla_{Median\_y}$ (Regression) | | |
|---|---|---|---|---|---|---|
| | Median | Predicted | Diff | Median | Predicted | Diff |
| Dataset1 | 6.3907 | 4.7771 | 25% | 3.6729 | 2.7773 | 24% |
| Dataset2 | 3.9209 | 3.2247 | 18% | 4.6643 | 3.4917 | 25% |
| Dataset3 | 6.041 | 4.8612 | 20% | 3.0294 | 2.6702 | 12% |
| Dataset4 | 9.2437 | 7.0653 | 24% | 3.5281 | 2.7442 | 22% |
| Dataset5 | 3.5112 | 2.8859 | 18% | 1.0073 | 0.8385 | 17% |
| Average | **21% ± 5%** | | | **20% ± 3%** | | |

The trained network prediction results for blocks with only one motion-compensated neighbor, whereas the Median vector is actually the value of the neighboring block's vector itself, are not as good as the Median vector and therefore we recommend using the base-line Median vector in this case.

*Table 13: MSE of regression Network-Predicted versus Median-Predicted for blocks with 2 motion compensated neighbors*

| 2 MVs | $1 - \nabla_{PMV_x}/\nabla_{Median\_x}$ (Regression) | | | $1 - \nabla_{PMV_y}/\nabla_{Median\_y}$ (Regression) | | |
|---|---|---|---|---|---|---|
| | Median | Predicted | Diff | Median | Predicted | Diff |
| Dataset1 | 12.9541 | 11.8424 | 9% | 6.805 | 6.4375 | 5% |
| Dataset2 | 14.2836 | 13.2947 | 7% | 7.1073 | 6.9341 | 2% |
| Dataset3 | 14.832 | 12.5091 | 16% | 6.4379 | 6.2164 | 3% |
| Dataset4 | 15.8313 | 11.3222 | 28% | 6.7581 | 5.1422 | 24% |
| Dataset5 | 14.7031 | 13.702 | 7% | 7.7121 | 7.3902 | 4% |
| Average | **13% ± 4%** | | | **8% ± 5%** | | |

We compared the results of the entropy and the number of coding bits to the results of the previous method – the classification network – when running the prediction for the FFW movies. The comparison results are depicted in **Table 14** and in **Table 15** for $MV_x$ and $MV_y$ respectively.

*Table 14: Entropy and number of Bits of the Median vector compared to the regression network predicted best PMV with respect to the GTMV. Calculated for Δx using 2 FFW movies.*

| | $Median_x$ | | $PMV_x$ (Regression) | | Improvement | |
|---|---|---|---|---|---|---|
| | Entropy | # Bits | Entropy | # Bits | Entropy | # Bits |
| Movie 1 | 1.821 | 5,617 | 1.0588 | 3,678 | 42% | 35% |
| Movie 2 | 2.3094 | 7,374 | 1.4136 | 4,615 | 39% | 37% |

As can be seen, the results of the regression network provide substantially better accuracy of the PMV for the FFW movies. This can be attributed to the relatively low classification accuracy that has been obtained while training the classification neural networks.

The number of layers used for the neural network has a substantial impact on the number of matrix multiplications required to predict MVs, thus reducing the number of layers is desirable in order to improve computational effectiveness. We have used different numbers of hidden layers within the range of $1 - 5$.

*Table 15: Entropy and number of Bits of the Median vector compared to the regression network predicted best PMV with respect to the GTMV. Calculated for Δy using 2 FFW movies.*

| | $Median_y$ | | $PMV_y$ (Regression) | | Improvement | |
|---|---|---|---|---|---|---|
| | Entropy | # Bits | Entropy | # Bits | Entropy | # Bits |
| Movie 1 | 1.096 | 3,779 | 0.6484 | 2,817 | 41% | 25% |
| Movie 2 | 1.6281 | 5,197 | 0.8824 | 3,273 | 46% | 37% |

Our results show that 1 hidden layer is sufficient to obtain the improved prediction results. Increasing the number of layers does not necessarily improve the prediction accuracy, as can be seen in **Table 16**. For each run we used 5 different datasets and averaged the improvement results. A plausible explanation is that the nature of the movements requires a function that can be satisfied with a 1-hidden-layer network. As we increase the number of layers unnecessarily, the number of redundant weights increases and given the same dataset size we reach overfitting and thus deteriorating accuracy results on the test dataset. In order to accomplish satisfactory training results with a larger number of layers, the dataset size has to be increased. More layers may be required when processing movies with higher motion activity.

*Table 16: MSE Prediction accuracy vs. the number of hidden layers*

| | $1 - \nabla_{PMV_x}/\nabla_{Median\_x}$ (Regression) | $1 - \nabla_{PMV_y}/\nabla_{Median\_y}$ (Regression) |
|---|---|---|
| 1 Hidden Layer | 26%±7% | 21%±4% |
| 2 Hidden Layers | 2%±6% | 18%±6% |
| 3 Hidden Layers | 5%±9% | 19%±5% |
| 4 Hidden Layers | 21%±3% | 20%±6% |
| 5 Hidden Layers | 22%±3% | 13%±5% |

### F. Neural Network Training Optimization

The networks were implemented in Python using Keras. The networks were trained with a full dataset per epoch. Keras criteria for halting training were defined with a patience parameter of 20 epochs of no improvement with a min-delta for loss function of 0.01. The FCNN used for classification and for regression are depicted in **Fig. 2** and in **Fig. 3** respectively. The networks were optimized using the Adam Optimizer [31], which is an improvement of Stochastic Gradient Decent (SGD) algorithm [32], that is using Momentum, which is effectively a factored running average of the gradients in the different steps so far, and RMSprop, which introduces a factored square of the gradient in order to reduce variations in steeper directions and prefer more gradual and stable ones. The Momentum, RMSprop and Adam optimizer formulas are described in the Appendix. We used the Adam optimizer with a constant learning rate of 0.001 and decay coefficients of 0.9 and 0.999 for $\beta_1$ and $\beta_2$ respectively, which are the default for Keras Adam optimizer library and were proven to provide sufficiently fast convergence rate.

### VII. CONCLUSIONS AND FUTURE WORK

In this paper we have proposed three algorithms to improve the value of the Predicted Motion Vector (PMV). We have demonstrated potential improvements in the order of 20% savings on the encoding efficiency of the PMV, when selecting the best PMV analytically. We further used a classification neural network for predicting which of the neighboring MVs is the best prediction of the GTMV. We have demonstrated a bit rate reduction between 5%-9% for fast forward movies which have high motion. The same applies to other movies with high motion whereas the magnitude of the MVs is large, such as movies taken from the cameras of autonomous vehicles, drones, and PTZ cameras. We further used a Fully Connected regression Neural Network (FCNN) approach for predicting the PMV for a target block from its neighboring motion compensated blocks. We have accomplished substantial accuracy improvement compared to the commonly used Median-based prediction. Our classification network for selecting the best neighboring PMV was not sufficiently accurate to reduce the bit rates of standard movies. However, it does reduce the bit rates of FFW movies by ~7%. While the analytic calculation of the best PMV from the 3 neighboring blocks provides the largest reduction of bit rate (~34%) for all movies, having to add signaling reduces the bits saving to ~20%. A regression network accomplishes the best improvement of FFW movies, reducing the bit rate to ~34%. The results of bit reduction for the 3 methods when applied to the FFW movies are depicted in **Table 17**.

*Table 17: Bits reduction comparison of the 3 proposed PMV methods*

|  | For $PMV_x$ | For $PMV_y$ |
|---|---|---|
| Best PMV | 20% | 11% |
| Classification | 9% | 5% |
| Regression | 36% | 31% |

The accuracy of the predicted PMVs can be further improved by incorporating additional, 2[nd] order neighboring motion compensated MVs and/or inter-frame MVs, taking advantage of the temporal domain. The same added 2[nd] order neighbors can also be used to improve the accuracy of the classification network and thus provide a closer prediction of the best neighbor MV, which clearly accomplishes large efficiency improvement. The improvements in the accuracy of the PMV can also be leveraged for exploring more computational efficient motion estimation algorithms. We are also proposing to improve the entropy of the predicted MVs and therefore coding efficiency, by incorporating a related criterion in the neural network loss function during training. In this paper we have not considered computation complexity. While the Best PMV method retains the same calculation complexity of the Median, the Classification and Regression neural networks increase the calculation complexity. Assuming that training is performed in advance and runtime only perform parallel matrix manipulations using Graphical Processing Units (GPU), we assume that the added complexity is acceptable and will still allow real time processing. However, this matter can be further explored and investigated in future research work.

## APPENDIX

The Momentum, RMSprop and Adam optimizer formulas are provided in equations (14), (15) and (16) respectively.

$$v_{t+1} = \rho v_t + g_t$$
$$w_{t+1} = w_t - \alpha v_{t+1} \quad (14)$$

Where for iteration time $t$,

$w_t$ corresponds to the weights that are updated

$v_t$ corresponds to the derivative of the gradient

$g_t$ corresponds to the gradient

$\rho$ is a friction hyperparameter momentum coefficient; and

$\alpha$ is the learning rate hyperparameter

$$m_{t+1}^2 = \beta m_t^2 + (1-\beta)g_t^2$$
$$w_{t+1} = w_t - \alpha g_t/(\sqrt{m_t^2} + \varepsilon) \quad (15)$$

Where for iteration time t,

$w_t$ corresponds to the weights that are updated

$g_t$ corresponds to the gradient

$m_t^2$ corresponds to a moving estimate of the squared gradient

$\beta$ is a decay rate hyperparameter

$\alpha$ is the learning rate hyperparameter; and

$\varepsilon$ is a small value that protects from dividing by zero

$$m_{t+1}^1 = \beta_1 m_t^1 + (1-\beta_1)g_t$$
$$m_{t+1}^2 = \beta_2 m_t^2 + (1-\beta_2)g_t^2$$
$$w_{t+1} = w_t - \alpha \widehat{m_t^1}/(\sqrt{\widehat{m_t^2}} + \varepsilon)$$
$$\widehat{m_t^1} = \frac{m_t^1}{(1-\beta_1^t)} \quad (16)$$
$$\widehat{m_t^2} = \frac{m_t^2}{(1-\beta_2^t)}$$

Where for iteration time t,

$w_t$ corresponds to the weights that are updated

$g_t$ corresponds to the gradient

$m_t^1$ corresponds to a moving estimate of the gradient

$m_t^2$ corresponds to a moving estimate of the squared gradient

$\beta_1$ and $\beta_2$ are decay rate hyperparameters

$\alpha$ is the learning rate hyperparameter; and

$\varepsilon$ is a small value that protects from dividing by zero.

ACKNOWLEDGMENT

This research work was partially supported by the Chateaubriand grant.